\documentclass[sn-aps,iicol]{sn-jnl}

\usepackage{graphicx}%
\usepackage{tabularx}%
\usepackage{multirow}%
\usepackage{amsmath,amssymb,amsfonts}%
\usepackage{amsthm}%
\usepackage{mathrsfs}%
\usepackage[title]{appendix}%
\usepackage{xcolor}%
\usepackage{textcomp}%
\usepackage{manyfoot}%
\usepackage{booktabs}%
\usepackage{algorithm}%
\usepackage{algorithmicx}%
\usepackage{algpseudocode}%
\usepackage{listings}%
\newcommand{\ham}{\widehat{\mathcal{H}}}
\newcommand{\meff}{\mathcal{M}}

\newcommand{\veff}{\mathcal{V}}

\newcommand{\ehfb}{E_{\textup{HFB}}}
\newcommand{\erot}{E_{\textup{rot}}}
\newcommand{\zpe}{E_{\textup{zpe}}}
\newcommand{\barn}{\text{b}}
\newcommand{\tsf}{t_{1/2}}
\newcommand{\eqc}[1]{Eq.~\eqref{#1}}
\newcommand{\figc}[1]{Fig.~\ref{#1}}

\newcommand{\secc}[1]{Sec.~\ref{#1}}
\newcommand{\ra}{$^{224}$Ra}
\newcommand{\pu}{$^{238}$Pu}

\newcommand{\pb}{$^{208}$Pb}
\graphicspath{{./}{figures/}}

\theoremstyle{thmstyleone}%
\theoremstyle{thmstyletwo}%

\theoremstyle{thmstylethree}%

\begin{document}

\title{Cluster properties of heavy nuclei predicted with the
Barcelona-Catania-Paris-Madrid energy density functional}


\author*[1,2]{\fnm{Samuel~A.} \sur{Giuliani}}\email{samuel.giuliani@uam.es}

\author*[1,3]{\fnm{Luis~M.} \sur{Robledo}}\email{luis.robledo@uam.es}


\affil*[1]{\orgdiv{Departamento de F{\'i}sica Te{\'o}rica and CIAFF}, \orgname{Universidad
Aut{\'o}noma de Madrid}, \orgaddress{\street{c.~Francisco Tom{\'a}s y Valiente
7}, \city{Cantoblanco}, \postcode{28049}, \state{Madrid}, \country{Spain}}}

\affil[2]{\orgdiv{Department of Physics, Faculty of Engineering and Physical
Sciences}, \orgname{University of Surrey}, \city{Guildford},
\postcode{GU2 7XH}, \country{United Kingdom}}

\affil[3]{\orgdiv{Center for Computational Simulation}, \orgname{Universidad
Polit{\'e}cnica de Madrid}, \orgaddress{\street{Campus Montegancedo},
\city{Boadilla del Monte}, \postcode{28660}, \state{Madrid}, \country{Spain}}}



\abstract{We study the cluster emission properties of \ra\ and \pu\ employing
	the Barcelona-Catania-Paris-Madrid (BCPM) energy density functional
	(EDF).  Starting from two-dimensional potential energy surfaces,
	coexisting fission paths are identified. A fission valley located at
	large octupole deformations, corresponding to a highly-asymmetric mass
	distribution, is found in both nuclei. As the corresponding fragments are
	dominated by the presence of \pb, we can relate this fission path to
	the emergence of cluster emission. Using the octupole moment as collective
	degree of freedom, we compute the cluster decay half-lives and study
	the impact of collective inertias, pairing strength and collective
	zero-point energy. The agreement with experimental data resembles the
results obtained for spontaneous fission half-lives, indicating the capability
of BCPM to consistently describe a large variety of fission phenomena,
including cluster emission.}




\maketitle

\section{Introduction}\label{sec:intro}
The phenomenon of cluster decay, or cluster radioactivity, consists in  the
spontaneous emission of fragments with an intermediate mass between fission and
$\alpha$ decay.  Originally postulated by Sandulescu \emph{et al.} in
1980~\cite{Sandulescu1980} and experimentally observed by Rose and Jones four
years later~\cite{Rose1984}, this decay is driven by the emission of one
fragment in the neighborhoods of the doubly magic nucleus \pb. From a
theoretical point of view, cluster radioactivity can be described as a
highly-asymmetric fission mode, where the fragments emerge as the consequence
of the quantum mechanical tunneling through the potential barrier of the
compound nucleus. Given the extreme mass asymmetries characterizing this decay,
the odd-parity octupole moment can be used as driving coordinate. Therefore,
cluster emission studies explore octupole deformation not only in the
traditional ground state regime, but also in the extreme
elongations required for the formation of a well defined neck that defines
scission point and fragment emission properties. Due to this, cluster decay has
been the subject of several experimental
measurements~\cite{Barwick1985,Price1985a,Wang1987,Wang1989,Price1989,Bonetti2001,Bonetti2007,Guglielmetti2008} and
microscopic studies based on effective nucleon-nucleon
interactions~\cite{Bhagwat2005,Xu2006,Warda2011,Adel2017,Zhao2023}, with a
renewed interest in the latest years arising from its predicted presence in the
super-heavy landscape~\cite{Warda2018,Matheson2019,Giuliani2019}.  

The Barcelona Catania Paris Madrid (BCPM)~\cite{Baldo2013} functional was
originally developed to describe with good accuracy binding energies and radii
of finite nuclei. Large scale mean-field Hartree-Fock-Bogoliubov (HFB)
calculations revealed that deformation properties of the functional are at the
level of other interactions such as Gogny D1S or D1M. This property opens up the
door to the use of BCPM in fission studies, and more specifically to describe
spontaneous fission lifetimes~\cite{Giuliani2013,Giuliani2014,Giuliani2018}.
One of the most useful features of BCPM is its relatively low computational
cost, specially when compared with finite range interactions like Gogny. This
feature makes possible large-scale fission calculations like the ones required
in stellar nucleosynthesis studies, including fission recycling of super-heavy
neutron rich nuclei~\cite{Giuliani2020}.

Given the success of BCPM in traditional fission studies, it seems pertinent to
explore the ability of the functional to describe very asymmetric fission modes
like the one taking place in cluster emission in actinides. This decay provides
an optimal case study for assessing the capability of microscopic models to
properly reproduce shell structure properties of parent and daughters nuclei at
large deformations, as well as the collective inertias required in the
calculation of cluster decay half-lives. In this work, we present the
results obtained for two different nuclei: $^{224}$Ra and $^{238}$Pu. The
former is one of the first cluster emitters detected in
laboratory~\cite{Price1985a},
and represents a canonical standard for studies concerning cluster emission. The
latter is a well known case of cluster emission observed in the transactinide
region~\cite{Wang1989}, and allows the exploration of cluster phenomena in heavier systems.
In this work, the cluster path is identified by computing the evolution of
the energy as a function of quadrupole and octupole deformation. Cluster
emission lifetimes are thus obtained using the octupole moment as collective
degree of freedom. Despite the sensitivity of the lifetimes with respect to the
different quantities entering the collective action
integral, a good agreement with experimental data is found. 

The paper is structured as follows. In Section~\ref{sec:meth}, the theoretical
framework employed in the calculation of the cluster decay half-lives is
outlined. In Section~\ref{sec:res}, fission pathways and cluster decay
half-lives predicted by BCPM are discussed. Conclusions are summarized in
Section~\ref{sec:conc}, together with an outlook on future work. 

\section{Methods}\label{sec:meth}
The self-consistent description of highly asymmetric fission is based on the 
HFB method with constraining operators~\cite{Schunck2016,Schunck2022}. In the HFB
method, the many-body wavefunction $|\Psi\rangle$ is defined as the vacuum to
all the annihilation quasiparticle operators $\beta_{\mu}$.
Nuclear states are thus obtained by minimizing the Routhian with constraints on the
neutron $N$ and proton $Z$ particle number, and the mass multipole moments $Q_{\mu\nu}$
\begin{equation}
	\ham' = \ham - \sum_{\tau=N,Z} \lambda_\tau \widehat{N}_\tau
	- \sum_{\mu=1,2,3} \lambda_{\mu}
	\widehat{Q}_{\mu0} \,,
	\label{eq:ehfb}
\end{equation}
being $\ham$ the HFB Hamiltonian. The independent $\lambda_i$ quantities are
determined by the condition that the gradient of the Routhian has to be
orthogonal to the gradient of the constraints. In this work, we impose
constraints on the axially symmetric quadrupole ($Q_{20}$), and
octupole ($Q_{30}$) moments, defined as
\begin{subequations}
	\begin{align}
		\widehat{Q}_{20} &= \hat{z}^2 - 
		\frac{1}{2}(\hat{x}^2+\hat{y}^2) \,; \\
		\widehat{Q}_{30} &= \hat{z}^3 - 
		\frac{3}{2}(\hat{x}^2+\hat{y}^2)\hat{z} \,.
	\label{eq:mult}
	\end{align}
\end{subequations}
Higher multipolarities are self-consistently adjusted in order to minimize the
total energy, e.g.\ the hexadecapole ($Q_{40}$) moment related to the
development of a fission neck:
\begin{align}
	\widehat{Q}_{40} &= \hat{z}^4 - 
	3(\hat{x}^2+\hat{y}^2)\hat{z}^2 +
	\frac{3}{8}(\hat{x}^2+\hat{y}^2)^2 \,.
\label{eq:q40}
\end{align}
The dipole moment $Q_{10}$ is set to zero to
avoid spuriousness associated with the centre-of-mass motion.

The HFB equations are solved using the HFBaxial computer code, which employs an
approximate second-order gradient method to determine the amplitudes of the
Bogoliubov transformation to quasiparticles~\cite{Robledo2011d}. The
quasiparticle operators are expanded in a large deformed axially-symmetric
harmonic oscillator (HO) basis, containing states with $J_z$ up to $35/2$ and
26 quanta along the $z$-direction. The HO quantum numbers are restricted by the
condition $n_z/q + 2n_\perp + |m| \leq N_0$, with $N_0=17$ and $q=1.5$. This
configuration is well suited for describing elongated shapes along the $z$
direction, as those routinely found in fission. The two oscillator lengths
characterising the HO bases are chosen as to minimize the energy for each
constrained configuration considered. 

One of the goals of the present paper is to evaluate the half-lives of spontaneous
cluster emission, which can be obtained by means of the Wenzel-Kramers-Brillouin
(WKB) method:
\begin{equation}
	t_{1/2} (s) = 2.86 \times 10^{-21} [1 + \exp(2S)]\,, 
	\label{eq:t12}
\end{equation}
being $S$ the collective action computed along the fission path
$L(Q_{30})$~\cite{Brack1972}
\begin{equation}
	S = \frac{1}{\hbar} \int_a^b dQ_{30} \sqrt{2\meff(Q_{30})
	(\veff(Q_{30}) - E_0)}  \,.
	\label{eq:action}
\end{equation}
In this work, collective inertias $\meff(Q_{30})$ are computed using the
perturbative cranking approximation within the adiabatic time-dependent HFB
(ATDHFB)~\cite{Girod1979} and the Gaussian Overlap Approximation to the
Generator Coordinate Method (GOA-GCM)~\cite{Ring1980} formalisms:
\begin{subequations}\label{eq:mcrank}
	\begin{align}
		\meff_\textup{ATDHFB} (Q_{30}) = \hbar^{2}
		\frac{M_{-3}(Q_{30})}{2M_{-1}^{2}(Q_{30})} \,;
		\label{eq:matd} \\
		\meff_\textup{GOA-GCM} (Q_{30}) = \hbar^{2}
		\frac{(M_{-2}(Q_{30}))^{2}}{2M_{-1}^{3}(Q_{30})} \,,
		\label{eq:mgcm}
	\end{align}
\end{subequations}
being $M_{-n}$ the energy-weighted moments obtained from the one-quasiparticle
energies $E_{\mu}$ and the two-quasiparticle zero-hole component of the octupole
operator $(Q^{20}_{30})_{\mu\nu}$ (see Appendix~E in Ref.~\cite{Ring1980}):
\begin{equation}
	M_{-n} (Q_{30}) = \sum_{\mu\nu}
	\frac{|\bigl(Q^{20}_{30}\bigr)_{\mu\nu}|^{2}}{(E_{\mu} -
	E_{\nu})^n} \,.
	\label{eq:mom}
\end{equation}
The effective potential $\veff(Q_{30})$ is obtained by subracting the rotational
and zero-point energy corrections from the HFB energy
\begin{equation}
	\veff(Q_{30})  = \ehfb(Q_{30}) - \erot(Q_{30}) - \zpe(Q_{30}) \,.
	\label{eq:veff}
\end{equation}
In this expression, $\erot$ represents the energy gained by restoring rotational
symmetry, which can be
effectively computed employing the recipe from Ref.~\cite{Egido2004}. $\zpe$ is the
energy correction associated to quantum fluctuations in the collective degree
of freedom $Q_{30}$, which is computed consistently with the collective inertia
scheme~\cite{Reinhard1975,Ring1980}:
\begin{subequations}\label{eq:zpe}
	\begin{align}
		\zpe^\textup{ATDHFB} (Q_{30}) =
		\frac{G(Q_{30})}{2\meff^{-1}_\textup{ATDHFB}(Q_{30})} \,; 
		\label{eq:zpeatd} \\
		\zpe^\textup{GOA-GCM} (Q_{30}) =
		\frac{G(Q_{30})}{2\meff^{-1}_\textup{GOA-GCM}(Q_{30})} \,, 
		\label{eq:zpegcm}
	\end{align}
\end{subequations}
being the overlaps 
\begin{equation}
	G(Q_{30}) = \frac{M_{-2}(Q_{30})}{2M_{-1}^{2}(Q_{30})} \,.
	\label{eq:overl}
\end{equation}
Finally, the action integral defined in \eqc{eq:action} is computed between the
classical turning points $a$ and $b$ determined by the condition $\veff = E_0$,
where $E_0$ represents the ground-state value of $\veff(Q_{30})$ plus the zero-point energy associated to the quantum motion
of the nucleus along the collective degrees of freedom. This quantity is
usually treated as a free parameter in order to reproduce spontaneous fission
lifetimes. In this work, we will study the impact of $E_0$ in the predicted
$\tsf$ by varying its value between 1.0 and 1.5~MeV.

To compute the effective potential energy $\veff(Q_{30})$ and collective
inertias $\meff(Q_{30})$ entering in \eqc{eq:action}, we employ the BCPM EDF\@.
For a comprehensive discussion regarding the details and applications
of this interaction, we refer to the recent review from
Baldo~\emph{et~al.}~\cite{Baldo2023}. Here, we just recap the basic ingredients
of the functional:
\begin{itemize}
\item A bulk or volume term is given by two fifth order polynomials of the 
proton and neutron densities. One of the polynomial is fitted to reproduce
the binding energy per nucleon in symmetric nuclear matter, while the other
uses the pure neutron matter equation of state. For intermediate situations,
a quadratic term proportional to  $\beta=\rho_{n}-\rho_{p}$ is used to
connect both polynomials.
\item A surface term, obtained from the Hartree potential of a finite 
range Gaussian.
\item A spin-orbit potential adopting the same form as the ones of Gogny or
Skyrme interactions, with a standard spin-orbit strength adapted to the
effective mass of BCPM being equal to the bare mass.
\item A Coulomb term, with two contributions, one the traditional Hartree
term derived from the Coulomb potential and the other the Slater local
approximation to Coulomb exchange.
\end{itemize}
The kinetic energy is computed using the standard quantum mechanic 
expression. 

\section{Results}\label{sec:res} 
In order to assess the capability of the BCPM interaction to reproduce the main
features of cluster decay in heavy nuclei, we follow the strategy of
Ref.~\cite{Warda2011} and study the properties of two representative nuclei:
\ra\ and \pu.

\subsection{Potential energy surfaces and fission paths}\label{sec:PES}
\begin{figure}[tb]
	\includegraphics[width=\columnwidth]{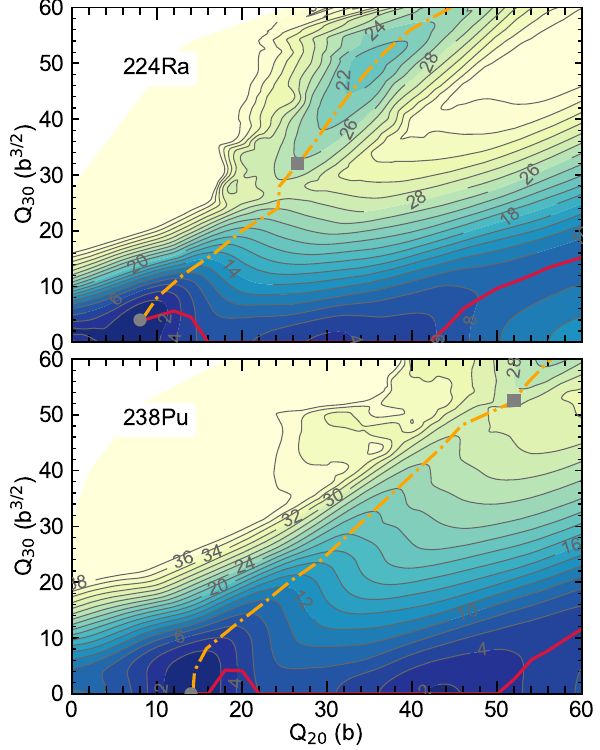}%
	\caption{Evolution of the HFB energy (including rotational correction)
		in MeV for \ra\ (upper panel) and \pu\ (lower panel), as a
		function of elongation ($Q_{20}$) and mass asymmetry
		($Q_{30}$). The ground state configuration of each nucleus is
		marked with a ${\color{gray}\bullet}$ symbol. The red solid
		line represents the traditional elongated fission path. The
	cluster path is marked with an orange dash-dotted line.\label{fig:pes2D}}
\end{figure}

\figc{fig:pes2D} shows the potential energy surface (PES) as a function of the
quadrupole $Q_{20}$ and octupole $Q_{30}$ mass moments predicted by BCPM for
\ra\ and \pu. This plot, representing the evolution of the HFB energy
(including the rotational correction) as a function of elongation and mass
asymmetry, allows for the identification of the multiple fission paths
coexisting in a particular nucleus.
Calculations are performed in a grid with a spacing of $2\,\barn$ in the $Q_{20}$
direction and $2\,\barn^{3/2}$ in the $Q_{30}$ direction. The harmonic oscillator
lengths have been optimized for each configuration. For a better visualization,
the predicted ground-state energy has been subtracted from the PES. 

\begin{figure}[tb]
	\includegraphics[width=\columnwidth]{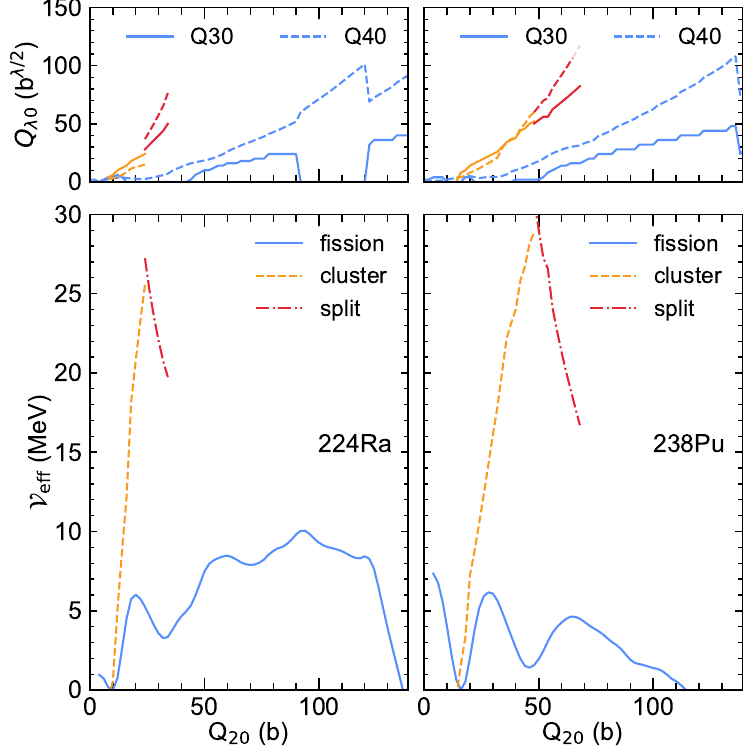}%
	\caption{Fission paths for \ra\ (left panels) and \pu\ (right panels) as
		a function of elongation. Lower panels: elongated
		fission path (blue solid lines), cluster path (orange dashed
		line), and total energy of the emerging cluster fragments (red
		dash-dotted line), in MeV. Upper panels: evolution of $Q_{30}$
		(solid lines) and $Q_{40}$ (dashed lines) mass multipole moments.
		\label{fig:pathQ20}}
\end{figure}

We observe that for both nuclei, BCPM predicts a well deformed ground-state
configuration: $Q^\textup{gs}_{20}=8\,$b for \ra, and
$Q^\textup{gs}_{20}=14\,$b for \pu. In the case of \ra\, the ground-state is
also predicted to break reflection symmetry as
$Q^\textup{gs}_{30}=4\,$b$^{3/2}$.  Starting from these ground-state
configuration, fission paths can be determined by seeking for the maximal
decrease in energy. Hence, we find that BCPM predicts the existence of two
distinct fission paths. The first one initially proceeds along configurations
with small $Q_{30}$ values.  After crossing the fission isomer located at
$Q_{20}^\textup{iso}=30\,\barn-44\,\barn$, it develops a strong octupole
deformation, finally reaching the outer turning point at $(Q^\textup{out}_{20},
Q^\textup{out}_{30})=(136\,\barn, 40\,\barn^{3/2})$, and $(Q^\textup{out}_{20},
Q^\textup{out}_{30})=(116\,\barn, 36\,\barn^{3/2})$ for \ra\ and \pu\,
respectively. This path represents a ``traditional'' elongated fission mode,
where the nucleus splits in two asymmetric fragments, giving raise to the
double-humped mass and charge fragment distributions characteristic of several
nuclei across the nuclear chart. The blue solid line in \figc{fig:pathQ20}
shows the one-dimensional projection of such fission path as a function of the
quadrupole moment, together with the evolution of $Q_{30}$, $Q_{40}$.
Both nuclei show an increase of hexadecapole deformation $Q_{40}$ with
increasing $Q_{20}$, suggesting the development of two fragments connected by a
thin fission neck, as typically found in elongated fission modes.  The
predicted inner barrier height is $6.73$~MeV for \ra, and $6.79$~MeV for \pu.
We also observe that the elongated fission path of \ra\ is particularly wide
and has a fission barrier height of $10.76$~MeV, indicating that BCPM
predicts \ra\ to be stable against fission. 

Besides this elongated asymmetric fission path, both PESs in \figc{fig:pes2D}
show the existence of a second fission valley located at very large octupole
deformations: $(Q_{20},Q_{30}=34\,\barn, 48\,\barn^{3/2})$ for \ra; and
$(Q_{20},Q_{30})=(66\,\text{b}, 76\,\text{b}^{3/2}$) for \pu. The octupole
moment along the fission path connecting this valley to the ground-state
configuration quickly grows with increasing elongation, suggesting the
emergence of a highly-asymmetric fission mode. The one dimensional projection
of this fission path for both nuclei is plotted in \figc{fig:pathQ20} as a
dashed red line. We notice that a saddle point is reached at $Q_{20}=26\,\barn$
for \ra, and $Q_{20}=52$ for \pu. At these deformations, the nucleus starts to
split into two fragments. For both \ra\ and \pu\, the heavy partner is a
spherical nucleus in the vicinity of $^{208}$Pb, clearly indicating that the
highly-asymmetric fission path found in these nuclei corresponds to a cluster
emission. We notice that the energy barrier of cluster decay is above 26~MeV,
indicating that the probability of the nucleus to undergo this decay is much
smaller than the one corresponding to the traditional fission decay. 

\begin{figure}[tb]
	\includegraphics[width=\columnwidth]{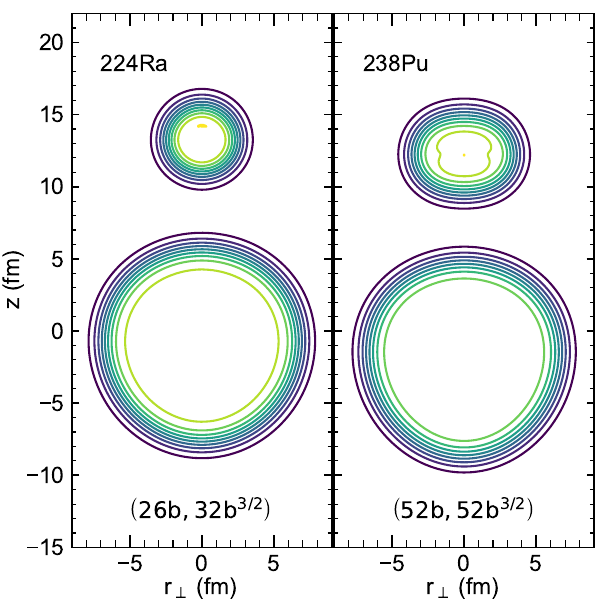}%
	\caption{Spatial mass distribution of \ra\ (left panel) and \pu\ (right
		panel). The configurations correspond to the
		${\color{gray}\blacksquare}$ symbols
		plotted in \figc{fig:pes2D}.\label{fig:dens}}
\end{figure}

In order to visualize the emerging fragments, \figc{fig:dens} shows the mass
distributions of \ra\ and \pu\ at the saddle point of the cluster path.
The cluster decay predicted by BCPM for these nuclei are $^{224}\text{Ra} \to
^{210}\text{Pb} + ^{14}\text{C}$ and $^{238}\text{Pu} \to ^{208}\text{Pb} +
^{30}\text{Mg}$, in agreement with experimental
observations~\cite{Bonetti2007}. In both nuclei, fragments' deformation are
determined by their ground-state deformation: spherical for $^{14}$C,
$^{208}$Pb and $^{210}$Pb, and oblate for $^{30}$Mg ($\beta_2=-0.17$). After
overcoming the saddle point of the cluster barrier, i.e., after the splitting
of the nucleus into two fragments, the energy of the nucleus starts to decrease
with elongation in a parabolic fashion. This is the result of the diminishing
Coulomb repulsion between the fragments as they move further apart. The outer
turning point of the cluster decay can thus be obtained by following the
Coulomb barrier until it falls below the ground-state energy. However, as
already discussed by Warda and Robledo~\cite{Warda2011}, we find that this
two-fragments configuration quickly saturates the HO basis for large values of
the elongation, suggesting that a larger basis should be employed in order to
compute the full cluster decay path.  Unfortunately, the number of HO shells is
limited by numerical accuracy, and cannot be increased without the emergence of
numerical instabilities in the estimation of the matrix elements. This limitation 
will be addressed in the next section.

\subsection{Cluster decay path as a function of $Q_{30}$}\label{sec:q3path}
As the cluster path overcomes the energy barrier, it proceeds towards larger
deformations by following the bottom of the highly-asymmetric valley described
in \secc{sec:PES}. As shown in \figc{fig:pes2D}, this results in a cluster path
placed along  a diagonal line in the $(Q_{20}, Q_{30})$ plane. Hence, a
correspondence can be established between increasing elongation and increasing
mass asymmetry for these cluster decays. As already proposed in
Ref.~\cite{Warda2011}, this proportionality can be exploited in order to
parametrize the cluster path as a function of the octupole moment $Q_{30}$.
This change of collective variable is particularly convient as the PES for a
fixed $Q_{30}$ is stiffer than for a fixed $Q_{20}$, which simplifies the
identification of the cluster decay path located at large $Q_{30}$ values.

\begin{figure}[tb]
	\includegraphics[width=\columnwidth]{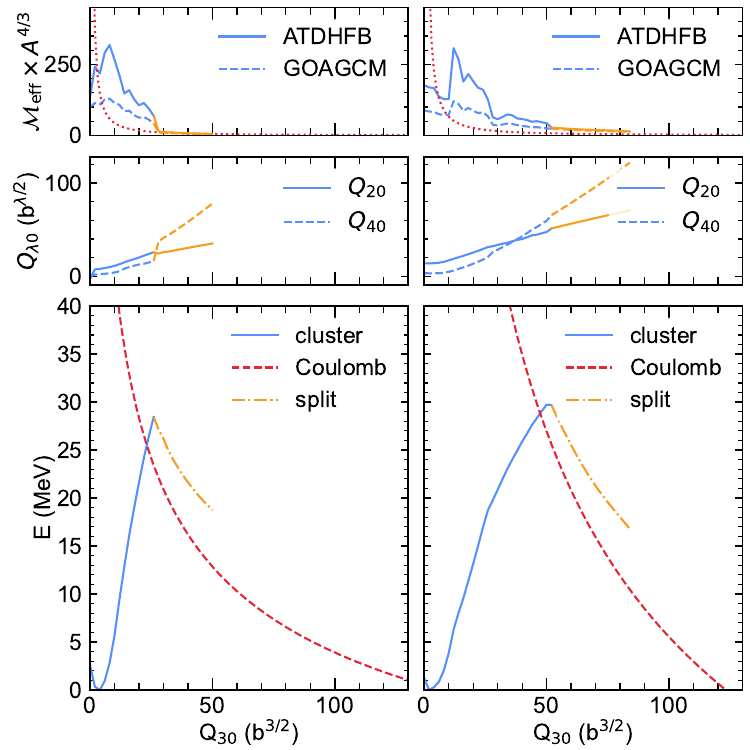}%
	\caption{Fission paths for \ra\ (left panels) and \pu\ (rigth panels) as
		a function of mass asymmetry $Q_{30}$. Lower panels: cluster
		path (blue solid line), energy of the emerging cluster
		fragments (orange dash-dotted line), and classical Coulomb
		repulsion energy (red dashed line), in MeV. Middle panels:
		evolution of $Q_{20}$ (solid lines) and $Q_{40}$ (dashed lines)
		mass multipole moments. Upper panels: ATDHFB~\eqref{eq:matd}
		(solid lines), GOA-GCM~\eqref{eq:mgcm} (dashed lines), and
		classical Coulomb~\eqref{eq:MQ3} (dotted line) 
		collective inertias in MeV$^{-1}$b$^{-3}$.
\label{fig:pathQ30}}
\end{figure}

The one-dimensional cluster path of \ra\ and \pu\ as a function of $Q_{30}$ is
plotted in \figc{fig:pathQ30}.
As in the $Q_{20}$ parametrization of \figc{fig:pathQ20}, we notice a fast
increase of the energy with increasing $Q_{30}$. A saddle point is reached at
$Q_{30}=26\,\barn^{3/2}$ for \ra\, and $Q_{30}=52\,\barn^{3/2}$ for \pu. At
these configurations, the nucleus is already split in two fragments,
corresponding to the cluster decay described in the previous section. Hence,
increasing the octupole moment beyond this point corresponds to a separation of
the centre of mass of the fragments, with the consequent diminishment of the
total energy due to the decreasing Coulomb repulsion. By approximating the
cluster fragments as two homogeneous spheres with charges $Z_1$ and $Z_2$, the
Coulomb repulsion energy as a function of the octupole moment can be written as
\begin{equation}
	V(Q_{30}) = V_\textup{C} - Q = e^2\frac{Z_1 Z_2}{R} - Q \,,
	\label{eq:vcoul}
\end{equation}
being $R$ the distance between the centers of mass of the fragments,
while the decay $Q$-value can be extracted from the AME2020
database~\cite{Wang2021}. The relationship between $R$ and the octupole moment
can be obtained by approximating the two fragments as spheres with masses
$A_1$ and $A_2$:
\begin{equation}
	Q_{30} = f_3 R^3\,,
	\label{eq:q30R}
\end{equation}
with
\begin{equation}
	f_3 = \frac{A_1 A_2}{A} \frac{|A_1-A_2|}{A}\,,
	\label{eq:f3}
\end{equation}
where $A$ is the mass number of the compound nucleus. The Coulomb repulsion
energy from \eqc{eq:vcoul} is plotted in \figc{fig:pathQ30} as a solid red
line. At the saddle point, the Coulomb repulsion and the HFB calculations
differ by roughly 5~MeV. This discrepancy is slightly larger than the one found
in Ref.~\cite{Warda2011}, and can be related to the excitation of the lighter
fragment in the presence of the Coulomb field produced and to the shape of the
emerging fragments. As a consequence of the saturation of the HO basis
discussed in \secc{sec:PES}, we notice a departure of the HFB calculations from
the Coulomb parabola with increasing $Q_{30}$. As this numerical limitation
imposed by the finite size of the basis do not allow a proper calculation of
configurations with large values of the octupole moment, in the next section
the cluster decay half-lives will be estimated employing the Coulomb barrier
from \eqc{eq:vcoul}. For this purpose, one shall rewrite the classical collective
inertia for two fragments separated by a distance $R$ in terms of the octupole
moment~\cite{Warda2011}:
\begin{equation}
	\meff(Q_{30}) = 
	\frac{\mu}{9 Q^{4/3}_{30} f_{3}^{2/3}} \,,
	\label{eq:MQ3}
\end{equation}
being $\mu=m_n A_1 A_2 / (A_1 + A_2)$ the reduced mass and $m_n$ the nucleon
mass. The upper panel of \figc{fig:pathQ30} shows the evolution of the
collective inertias from Eqs.~\eqref{eq:matd},~\eqref{eq:mgcm}
and~\eqref{eq:MQ3}. The microscopic ATDHFB and GOA-GCM inertias present large
fluctuations due to the crossing of single particle levels, which are obviously
absent in the classical collective inertias.  However, once the parental
nucleus splits into two fragments, both microscopic inertias agree to the same
asymptotic value.

\subsection{Cluster decay half-lives}
In order to estimate the spontaneous fission half-lives of cluster decay, we
employ the formalism described in \secc{sec:meth}. The fission path
$L(Q_{30})$ employed in
the estimation action integral is obtained by splicing the cluster path
(parametrized as a function of $Q_{30}$) with the Coulomb repulsion barrier
obtained in \secc{sec:q3path}. The transition between the two regimes occur at
the saddle point ($Q_{30}=22\,\barn^{3/2}$ for \ra, and
$Q_{30}=46\,\barn^{3/2}$ for \pu), keeping the collective inertias consistent
with the employed barrier. 

\begin{figure}[tb] \includegraphics[width=\columnwidth]{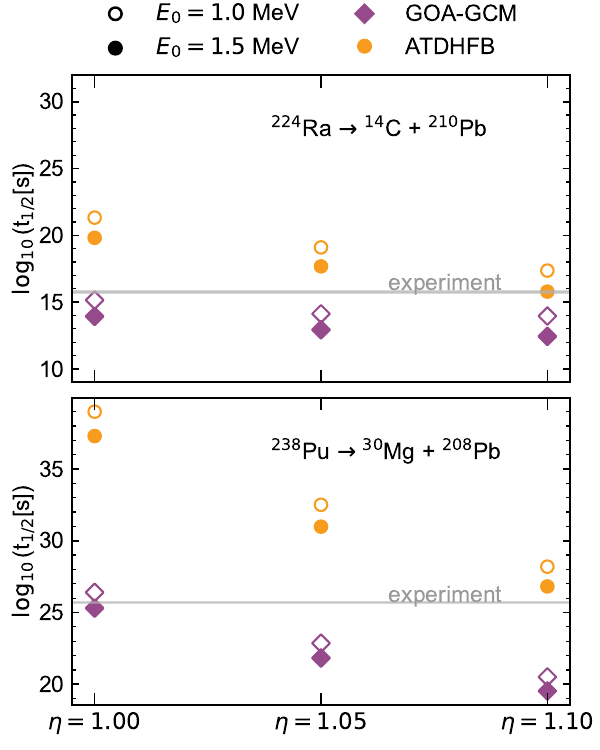}%
	\caption{Spontaneous fission half-lives for the cluster decay of \ra\
		(upper panel) and \pu\ (lower panel) as a function of different
		values of the pairing strength $\eta$. Results obtained with
		ATDHFB collective inertias~\eqref{eq:matd} and
		GOA-GCM~\eqref{eq:mgcm} are plotted as orange circles and
		purple diamonds, respectively. Empty and full symbols
		represents the half-lives computed with $E_0=1.0$ and~1.5,
		respectively. The horizontal gray band shows the experimental
data~\cite{Bonetti2007}.\label{fig:tsf}}
\end{figure}

The cluster-decay  half-lives, computed from \eqc{eq:action}, are plotted in
\figc{fig:tsf} for two different values of the collective zero-point energy
$E_0$.
The comparison with experimental data~\cite{Bonetti2007} shows a good agreement
when the GOA-GCM collective inertias are employed, while with the ATDHFB scheme
the half-lives are overestimated (particularly for \pu). In general, we notice
a large scatter in the predicted results depending on the collective inertia
scheme and $E_0$ value employed for the estimation of $\tsf$. This is not a
surprising result, as both nuclei have a high fission barrier (which in turn
implies a small cluster emission probability and long half-life), which makes
the theoretical calculations of $\tsf$ extremely sensitive to the details
concerning the estimation of the quantities entering into \eqc{eq:action}.
Regarding the absolute value of theoretical half-lives, it should be mentioned
that the inclusion of pairing as collective degree of freedom can reduce the
predicted $\tsf$ of long-lived nuclei by several orders of
magnitude~\cite{Moretto1974,Giuliani2014,Sadhukhan2014,Zhao2016}. This is
because collective inertias decrease as the square of the inverse pairing gap,
leading to a reduction of the $\tsf$ if the action integral in \eqc{eq:action}
is computed along the least-action path. While the inclusion of paring as
dynamical variable goes beyond the scope of this work, its effect on the
collective inertias and $\tsf$ can be mimicked by increasing the pairing
strength $\eta$~\cite{Giuliani2013,Rodriguez-Guzman2014,Rodriguez-Guzman2014a}.
\figc{fig:tsf} shows the predicted half-lives for cluster emission when pairing
strength is increased by 5\% ($\eta=1.05$) and 10\% ($\eta=1.10$). From this
plot it is possible to conclude that for larger $\eta$ values not only
the $\tsf$ are quenched, as expected, but also the spread in
theoretical predictions due to variations in $E_0$ and the collective inertia
scheme is reduced, in agreement with the results obtained when pairing is included as
collective degree of freedom~\cite{Giuliani2014,Sadhukhan2014,Zhao2016}.
Evidently, the choice of the collective inertia scheme is the main source of
uncertainty, due to the large values of the collective action integral of these
decays. While a better agreement with experimental data is
obtained with the GOA-GCM scheme, as these are systematically smaller than the
ATDHFB ones, an increase of the pairing strength by a 10\% brings the ATDHFB
predictions on top of experimental data, similarly to what already found in
BCPM fission studies in transactinides nuclei~\cite{Giuliani2013}. However, we
stress out again that theoretical $\tsf$ are extremely sensitive to the details
of the action integral calculation, and therefore the comparison with
experimental data should be always taken with a grain of salt. Nevertheless, we
conclude that BCPM provides a good reproduction of cluster decay half-lives
within the simplified scheme presented in this work.

\section{Conclusions}\label{sec:conc}
The cluster-decay properties of \ra\ and \pu\ have been computed using the BCPM
EDF\@. By studying the evolution of the potential energy surface as a function
of the mass quadrupole moment (elongation) and octupole moment (mass asymmetry)
we identified the coexisting fission paths in these nuclei. Besides the
traditional elongated fission mode, which proceeds through low values of the
octupole moment, BCPM predicts the existence of a highly-asymmetric fission
valley located at large $Q_{30}$, in agreement with previous EDFs
studies~\cite{Warda2011}. The large value of the cluster barrier indicates that
this mode is substantially suppressed compared to the traditional fission mode,
as observed in laboratory.

The cluster decay half-lives are computed using the WKB formalism, by
parametrizing the fission path as a function of $Q_{30}$. Due to the large
values of the fission barriers, we find a large impact of collective inertias
and ground-state zero-point energy on the estimated $\tsf$. By increasing the
pairing strength, the sensitivity of the estimated half-lives on the different
quantities defining the collective action diminishes. Overall, a good
agreement with experimental data is found when the GOA-GCM scheme is employed.
Conversely, for the ATDHFB inertias an increase in the pairing strength by
a 10\% is required in order to reproduce the observed $\tsf$.

In a future work, we plan to expand and improve the formalism employed for the
estimation of the half-lives. First and foremost, the impact of including
pairing as a dynamical variable should be assesed, since it could reduce the
predicted $\tsf$ by several orders of magnitude. Furthermore, the estimation
of the collective inertias can be improved by employing the non-perturbative
ATDHFB and GOA-GCM schemes~\cite{Yuldashbaeva1999,Baran2011,Giuliani2018b}.
Work along these lines is in progress and will be the subject of a future
publication.

\backmatter


\bmhead{Acknowledgments}
Some figures shown in this paper were created following the colorblind friendly
color scheme from Ref.~\cite{Petroff2021}. This work has been supported by the
Spanish Agencia Estatal de Investigaci{\'o}n (AEI) of the Ministry of Science
and Innovation (MCIN) under grant agreement No.~PID2021-127890NB-I00. SG
acknowledges support by the ``Ram{\'o}n y Cajal'' grant No.~RYC2021-031880-I
funded by MCIN/AEI/10.13039/501100011033  and the European
Union-``NextGenerationEU''.

\bibliography{library.bib}

\begin{thebibliography}{10}
\providecommand{\url}[1]{{#1}}
\providecommand{\urlprefix}{URL }
\providecommand{\doi}[1]{\url{https://doi.org/#1}}
\bibcommenthead

\bibitem{Sandulescu1980}
A.~Sandulescu, D.N. Poenaru, W.~Greiner, {New type of decay of heavy nuclei
  intermediate between fission and alpha decay}.
\newblock Sov. J. Part. Nucl. \textbf{11}(6) (1980).
\newblock \urlprefix\url{https://www.osti.gov/biblio/6189038}

\bibitem{Rose1984}
H.J. Rose, G.A. Jones, {A new kind of natural radioactivity}.
\newblock Nature \textbf{307}(5948), 245--247 (1984).
\newblock \doi{10.1038/307245a0}.
\newblock \urlprefix\url{https://www.nature.com/articles/307245a0}

\bibitem{Barwick1985}
S.W. Barwick, P.B. Price, J.D. Stevenson, {Radioactive decay of 232U by 24Ne
  emission}.
\newblock Phys. Rev. C \textbf{31}(5), 1984--1986 (1985).
\newblock \doi{10.1103/PhysRevC.31.1984}.
\newblock \urlprefix\url{https://link.aps.org/doi/10.1103/PhysRevC.31.1984}

\bibitem{Price1985a}
P.B. Price, J.D. Stevenson, S.W. Barwick, H.L. Ravn, {Discovery of Radioactive
  Decay of 222Ra and 224Ra by 14C}.
\newblock Phys. Rev. Lett. \textbf{54}(4), 297--299 (1985).
\newblock \doi{10.1103/PhysRevLett.54.297}.
\newblock \urlprefix\url{https://link.aps.org/doi/10.1103/PhysRevLett.54.297}

\bibitem{Wang1987}
S.~Wang, P.B. Price, S.W. Barwick, K.J. Moody, E.K. Hulet, {Radioactive decay
  of 234U via Ne and Mg emission}.
\newblock Phys. Rev. C \textbf{36}(6), 2717--2720 (1987).
\newblock \doi{10.1103/PhysRevC.36.2717}.
\newblock \urlprefix\url{https://link.aps.org/doi/10.1103/PhysRevC.36.2717}

\bibitem{Wang1989}
S.~Wang, D.~Snowden-Ifft, P.B. Price, K.J. Moody, E.K. Hulet, {Heavy fragment
  radioactivity of 238Pu: Si and Mg emission}.
\newblock Phys. Rev. C \textbf{39}(4), 1647--1650 (1989).
\newblock \doi{10.1103/PhysRevC.39.1647}.
\newblock \urlprefix\url{https://link.aps.org/doi/10.1103/PhysRevC.39.1647}

\bibitem{Price1989}
P.B. Price, {Heavy-Particle Radioactivity (A $>$ 4)}.
\newblock Annu. Rev. Nucl. Part. Sci. \textbf{39}(1), 19--42 (1989).
\newblock \doi{10.1146/annurev.ns.39.120189.000315}.
\newblock
  \urlprefix\url{https://www.annualreviews.org/doi/10.1146/annurev.ns.39.120189.000315}

\bibitem{Bonetti2001}
R.~Bonetti, C.~Carbonini, A.~Guglielmetti, M.~Hussonnois, D.~Trubert, C.~{Le
  Naour}, {Cluster decay of 230U via Ne emission}.
\newblock Nucl. Phys. A \textbf{686}(1-4), 64--70 (2001).
\newblock \doi{10.1016/S0375-9474(00)00508-X}.
\newblock
  \urlprefix\url{https://linkinghub.elsevier.com/retrieve/pii/S037594740000508X}

\bibitem{Bonetti2007}
R.~Bonetti, A.~Guglielmetti, {Cluster radioactivity: an overview after twenty
  years}.
\newblock Rom. Reports Phys. \textbf{59}(2), 301 (2007)

\bibitem{Guglielmetti2008}
A.~Guglielmetti, D.~Faccio, R.~Bonetti, S.V. Shishkin, S.P. Tretyakova, S.V.
  Dmitriev, A.A. Ogloblin, G.A. Pik-Pichak, N.P. van~der Meulen, G.F. Steyn,
  T.N. van~der Walt, C.~Vermeulen, D.~McGee, {Carbon radioactivity of 223 Ac
  and a search for nitrogen emission}.
\newblock J. Phys. Conf. Ser. \textbf{111}(1), 012050 (2008).
\newblock \doi{10.1088/1742-6596/111/1/012050}.
\newblock
  \urlprefix\url{https://iopscience.iop.org/article/10.1088/1742-6596/111/1/012050}

\bibitem{Bhagwat2005}
A.~Bhagwat, Y.K. Gambhir, {Relativistic mean field description of cluster
  radioactivity}.
\newblock Phys. Rev. C \textbf{71}(1), 017301 (2005).
\newblock \doi{10.1103/PhysRevC.71.017301}.
\newblock \urlprefix\url{https://link.aps.org/doi/10.1103/PhysRevC.71.017301}

\bibitem{Xu2006}
F.~Xu, J.~Pei, {Mean-field cluster potentials for various cluster decays}.
\newblock Phys. Lett. B \textbf{642}(4), 322--325 (2006).
\newblock \doi{10.1016/j.physletb.2006.09.048}.
\newblock
  \urlprefix\url{https://linkinghub.elsevier.com/retrieve/pii/S0370269306012275}

\bibitem{Warda2011}
M.~Warda, L.M. Robledo, {Microscopic description of cluster radioactivity in
  actinide nuclei}.
\newblock Phys. Rev. C \textbf{84}(4), 44608 (2011).
\newblock \doi{10.1103/PhysRevC.84.044608}.
\newblock \urlprefix\url{http://link.aps.org/doi/10.1103/PhysRevC.84.044608}.
\newblock {\href{https://arxiv.org/abs/1107.1478}{{arXiv:1107.1478}}}

\bibitem{Adel2017}
A.~Adel, T.~Alharbi, {Cluster decay half-lives of trans-lead nuclei based on a
  finite-range nucleon–nucleon interaction}.
\newblock Nucl. Phys. A \textbf{958}, 187--201 (2017).
\newblock \doi{10.1016/j.nuclphysa.2016.12.002}.
\newblock \urlprefix\url{http://dx.doi.org/10.1016/j.nuclphysa.2016.12.002
  https://linkinghub.elsevier.com/retrieve/pii/S0375947416302986}

\bibitem{Zhao2023}
J.~Zhao, J.P. Ebran, L.~Heitz, E.~Khan, F.~Mercier, T.~Nik{\v{s}}i{\'{c}},
  D.~Vretenar, {Microscopic description of $\alpha$, 2$\alpha$, and cluster
  decays of Rn 216-220 and Ra 220-224}.
\newblock Phys. Rev. C \textbf{107}(3), 1--9 (2023).
\newblock \doi{10.1103/PhysRevC.107.034311}

\bibitem{Warda2018}
M.~Warda, A.~Zdeb, L.M. Robledo, {Cluster radioactivity in superheavy nuclei}.
\newblock Phys. Rev. C \textbf{98}(4), 041602 (2018).
\newblock \doi{10.1103/PhysRevC.98.041602}.
\newblock \urlprefix\url{http://arxiv.org/abs/1807.00342
  https://link.aps.org/doi/10.1103/PhysRevC.98.041602}.
\newblock {\href{https://arxiv.org/abs/1807.00342}{{arXiv:1807.00342}}}

\bibitem{Matheson2019}
Z.~Matheson, S.A. Giuliani, W.~Nazarewicz, J.~Sadhukhan, N.~Schunck, {Cluster
  radioactivity of $^{294}_{118}$Og$_{176}$}.
\newblock Phys. Rev. C \textbf{99}(4), 041304 (2019).
\newblock \doi{10.1103/PhysRevC.99.041304}.
\newblock \urlprefix\url{https://link.aps.org/doi/10.1103/PhysRevC.99.041304}

\bibitem{Giuliani2019}
S.A. Giuliani, Z.~Matheson, W.~Nazarewicz, E.~Olsen, P.G. Reinhard,
  J.~Sadhukhan, B.~Schuetrumpf, N.~Schunck, P.~Schwerdtfeger, {Colloquium :
  Superheavy elements: Oganesson and beyond}.
\newblock Rev. Mod. Phys. \textbf{91}(1), 011001 (2019).
\newblock \doi{10.1103/RevModPhys.91.011001}.
\newblock \urlprefix\url{https://doi.org/10.1103/RevModPhys.91.011001
  https://link.aps.org/doi/10.1103/RevModPhys.91.011001}

\bibitem{Baldo2013}
M.~Baldo, L.M. Robledo, P.~Schuck, X.~Vi{\~{n}}as, {New Kohn-Sham density
  functional based on microscopic nuclear and neutron matter equations of
  state}.
\newblock Phys. Rev. C \textbf{87}(6), 064305 (2013).
\newblock \doi{10.1103/PhysRevC.87.064305}.
\newblock \urlprefix\url{http://arxiv.org/abs/1210.1321
  http://link.aps.org/doi/10.1103/PhysRevC.87.064305}.
\newblock {\href{https://arxiv.org/abs/1210.1321}{{arXiv:1210.1321}}}

\bibitem{Giuliani2013}
S.A. Giuliani, L.M. Robledo, {Fission properties of the
  Barcelona-Catania-Paris-Madrid energy density functional}.
\newblock Phys. Rev. C \textbf{88}(5), 054325 (2013).
\newblock \doi{10.1103/PhysRevC.88.054325}.
\newblock \urlprefix\url{https://link.aps.org/doi/10.1103/PhysRevC.88.054325}.
\newblock {\href{https://arxiv.org/abs/1305.0293}{{arXiv:1305.0293}}}
  {[nucl-th]}

\bibitem{Giuliani2014}
S.A. Giuliani, L.M. Robledo, R.R. Rodr{\'{i}}guez-Guzm{\'{a}}n, {Dynamic versus
  static fission paths with realistic interactions}.
\newblock Phys. Rev. C \textbf{90}(5), 054311 (2014).
\newblock \doi{10.1103/PhysRevC.90.054311}.
\newblock \urlprefix\url{https://link.aps.org/doi/10.1103/PhysRevC.90.054311}.
\newblock {\href{https://arxiv.org/abs/1312.7229}{{arXiv:1312.7229}}}

\bibitem{Giuliani2018}
S.A. Giuliani, G.~Mart{\'{i}}nez-Pinedo, L.M. Robledo, {Fission properties of
  superheavy nuclei for r-process calculations}.
\newblock Phys. Rev. C \textbf{97}(3), 034323 (2018).
\newblock \doi{10.1103/PhysRevC.97.034323}.
\newblock \urlprefix\url{https://link.aps.org/doi/10.1103/PhysRevC.97.034323}.
\newblock {\href{https://arxiv.org/abs/1704.00554}{{arXiv:1704.00554}}}

\bibitem{Giuliani2020}
S.A. Giuliani, G.~Mart{\'{i}}nez-Pinedo, M.R. Wu, L.M. Robledo, {Fission and
  the r-process nucleosynthesis of translead nuclei in neutron star mergers}.
\newblock Phys. Rev. C \textbf{102}(4), 045804 (2020).
\newblock \doi{10.1103/PhysRevC.102.045804}.
\newblock \urlprefix\url{http://arxiv.org/abs/1904.03733
  https://link.aps.org/doi/10.1103/PhysRevC.102.045804}.
\newblock {\href{https://arxiv.org/abs/1904.03733}{{arXiv:1904.03733}}}

\bibitem{Schunck2016}
N.~Schunck, L.M. Robledo, {Microscopic theory of nuclear fission: A review}.
\newblock Reports Prog. Phys. \textbf{79}(11), 116301 (2016).
\newblock \doi{10.1088/0034-4885/79/11/116301}.
\newblock
  \urlprefix\url{http://arxiv.org/abs/1511.07517%0Ahttp://dx.doi.org/10.1088/0034-4885/79/11/116301
  http://stacks.iop.org/0034-4885/79/i=11/a=116301?key=crossref.30cd21ca9f619239da027337924f470b}.
\newblock {\href{https://arxiv.org/abs/1511.07517}{{arXiv:1511.07517}}}

\bibitem{Schunck2022}
N.~Schunck, D.~Regnier, {Theory of nuclear fission}.
\newblock Prog. Part. Nucl. Phys. \textbf{125}, 103963 (2022).
\newblock \doi{10.1016/j.ppnp.2022.103963}.
\newblock \urlprefix\url{https://doi.org/10.1016/j.ppnp.2022.103963
  https://linkinghub.elsevier.com/retrieve/pii/S0146641022000242}.
\newblock {\href{https://arxiv.org/abs/2201.02719}{{arXiv:2201.02719}}}

\bibitem{Robledo2011d}
L.M. Robledo, G.F. Bertsch, {Application of the gradient method to
  Hartree-Fock-Bogoliubov theory}.
\newblock Phys. Rev. C \textbf{84}(1), 014312 (2011).
\newblock \doi{10.1103/PhysRevC.84.014312}.
\newblock \urlprefix\url{http://link.aps.org/doi/10.1103/PhysRevC.84.014312
  https://link.aps.org/doi/10.1103/PhysRevC.84.014312}.
\newblock {\href{https://arxiv.org/abs/1104.5453}{{arXiv:1104.5453}}}

\bibitem{Brack1972}
M.~Brack, J.~Damgaard, A.S. Jensen, H.C. Pauli, V.M. Strutinsky, C.Y. Wong,
  {Funny Hills: The Shell-Correction Approach to Nuclear Shell Effects and Its
  Applications to the Fission Process}.
\newblock Rev. Mod. Phys. \textbf{44}(2), 320--405 (1972).
\newblock \doi{10.1103/RevModPhys.44.320}.
\newblock \urlprefix\url{https://link.aps.org/doi/10.1103/RevModPhys.44.320}

\bibitem{Girod1979}
M.~Girod, B.~Grammaticos, {The zero-point energy correction and its effect on
  nuclear dynamics}.
\newblock Nucl. Phys. A \textbf{330}(1), 40--52 (1979).
\newblock \doi{10.1016/0375-9474(79)90535-9}

\bibitem{Ring1980}
P.~Ring, P.~Schuck, \emph{{The Nuclear Many-Body Problem}}, 1st edn. (Springer
  Berlin Heidelberg, Berlin, Heidelberg, 1980), pp. 1--718.
\newblock \urlprefix\url{http://www.springer.com/us/book/9783540212065}

\bibitem{Egido2004}
J.L. Egido, L.M. Robledo, in \emph{Ext. Density Funct. Nucl. Struct. Phys.},
  vol. 641, ed. by G.~Lalazissis, P.~Ring, D.~Vretenar (Springer Berlin
  Heidelberg, Berlin, Heidelberg, 2004), chap.~10, pp. 269--302.
\newblock \doi{10.1007/978-3-540-39911-7_10}.
\newblock \urlprefix\url{http://link.springer.com/10.1007/978-3-540-39911-7_10
  http://arxiv.org/abs/nucl-th/0311106}

\bibitem{Reinhard1975}
P.~Reinhard, {Zero-point energies in the two-center shell model}.
\newblock Nucl. Phys. A \textbf{252}(1), 133--140 (1975).
\newblock \doi{10.1016/0375-9474(75)90607-7}.
\newblock
  \urlprefix\url{https://linkinghub.elsevier.com/retrieve/pii/0375947475906077}

\bibitem{Baldo2023}
M.~Baldo, L.M. Robledo, X.~Vi{\~{n}}as, {The Barcelona Catania Paris Madrid
  energy density functional}.
\newblock Eur. Phys. J. A \textbf{59}(7), 156 (2023).
\newblock \doi{10.1140/epja/s10050-023-01062-z}.
\newblock \urlprefix\url{https://doi.org/10.1140/epja/s10050-023-01062-z
  https://link.springer.com/10.1140/epja/s10050-023-01062-z}

\bibitem{Wang2021}
M.~Wang, W.J. Huang, F.~Kondev, G.~Audi, S.~Naimi, {The AME 2020 atomic mass
  evaluation (II). Tables, graphs and references}.
\newblock Chinese Phys. C \textbf{45}(3), 030003 (2021).
\newblock \doi{10.1088/1674-1137/abddaf}.
\newblock
  \urlprefix\url{https://iopscience.iop.org/article/10.1088/1674-1137/abddaf}

\bibitem{Moretto1974}
L.G. Moretto, R.P. Babinet, {Large superfluidity enhancement in the penetration
  of the fission barrier}.
\newblock Phys. Lett. B \textbf{49}(2), 147--149 (1974).
\newblock \doi{10.1016/0370-2693(74)90494-8}.
\newblock
  \urlprefix\url{http://www.sciencedirect.com/science/article/pii/0370269374904948
  http://linkinghub.elsevier.com/retrieve/pii/0370269374904948}

\bibitem{Sadhukhan2014}
J.~Sadhukhan, J.~Dobaczewski, W.~Nazarewicz, J.A. Sheikh, A.~Baran,
  {Pairing-induced speedup of nuclear spontaneous fission}.
\newblock Phys. Rev. C \textbf{90}, 1--5 (2014).
\newblock \doi{10.1103/PhysRevC.90.061304}.
\newblock \urlprefix\url{http://link.aps.org/doi/10.1103/PhysRevC.90.061304}

\bibitem{Zhao2016}
J.~Zhao, B.N. Lu, T.~Nik{\v{s}}i{\'{c}}, D.~Vretenar, S.G. Zhou,
  {Multidimensionally-constrained relativistic mean-field study of spontaneous
  fission: Coupling between shape and pairing degrees of freedom}.
\newblock Phys. Rev. C \textbf{93}(4), 044315 (2016).
\newblock \doi{10.1103/PhysRevC.93.044315}.
\newblock \urlprefix\url{http://arxiv.org/abs/1603.00992
  http://link.aps.org/doi/10.1103/PhysRevC.93.044315}.
\newblock {\href{https://arxiv.org/abs/1603.00992}{{arXiv:1603.00992}}}

\bibitem{Rodriguez-Guzman2014}
R.R. Rodr{\'{i}}guez-Guzm{\'{a}}n, L.M. Robledo, {Microscopic description of
  fission in uranium isotopes with the Gogny energy density functional}.
\newblock Phys. Rev. C \textbf{89}(5), 054310 (2014).
\newblock \doi{10.1103/PhysRevC.89.054310}.
\newblock \urlprefix\url{https://link.aps.org/doi/10.1103/PhysRevC.89.054310}.
\newblock {\href{https://arxiv.org/abs/1312.7229}{{arXiv:1312.7229}}}

\bibitem{Rodriguez-Guzman2014a}
R.R. Rodr{\'{i}}guez-Guzm{\'{a}}n, L.M. Robledo, {Microscopic description of
  fission in neutron-rich plutonium isotopes with the Gogny-D1M energy density
  functional}.
\newblock Eur. Phys. J. A \textbf{50}(9), 142 (2014).
\newblock \doi{10.1140/epja/i2014-14142-6}.
\newblock \urlprefix\url{http://link.springer.com/10.1140/epja/i2014-14142-6}.
\newblock {\href{https://arxiv.org/abs/1405.6784}{{arXiv:1405.6784}}}

\bibitem{Yuldashbaeva1999}
E.~Yuldashbaeva, J.~Libert, P.~Quentin, M.~Girod, {Mass parameters for large
  amplitude collective motion: A perturbative microscopic approach}.
\newblock Phys. Lett. B \textbf{461}(1-2), 1--8 (1999).
\newblock \doi{10.1016/S0370-2693(99)00836-9}.
\newblock
  \urlprefix\url{http://www.sciencedirect.com/science/article/pii/S0370269399008369
  http://linkinghub.elsevier.com/retrieve/pii/S0370269399008369}

\bibitem{Baran2011}
A.~Baran, J.A. Sheikh, J.~Dobaczewski, W.~Nazarewicz, A.~Staszczak, {Quadrupole
  collective inertia in nuclear fission: Cranking approximation}.
\newblock Phys. Rev. C \textbf{84}(5), 054321 (2011).
\newblock \doi{10.1103/PhysRevC.84.054321}.
\newblock \urlprefix\url{https://link.aps.org/doi/10.1103/PhysRevC.84.054321}.
\newblock {\href{https://arxiv.org/abs/1007.3763}{{arXiv:1007.3763}}}

\bibitem{Giuliani2018b}
S.A. Giuliani, L.M. Robledo, {Non-perturbative collective inertias for fission:
  A comparative study}.
\newblock Phys. Lett. B \textbf{787}, 134--140 (2018).
\newblock \doi{10.1016/j.physletb.2018.10.045}.
\newblock
  \urlprefix\url{https://linkinghub.elsevier.com/retrieve/pii/S0370269318308165
  http://www.sciencedirect.com/science/article/pii/S0370269318308165}

\bibitem{Petroff2021}
M.A. Petroff, {Accessible Color Sequences for Data Visualization}  (2021).
\newblock \urlprefix\url{http://arxiv.org/abs/2107.02270}.
\newblock {\href{https://arxiv.org/abs/2107.02270}{{arXiv:2107.02270}}}

\end{thebibliography}

\end{document}